\documentclass[a4paper,12pt]{spieman}  

\usepackage{amsmath,amsfonts,amssymb}
\usepackage{setspace}
\usepackage{tocloft}
\usepackage{textgreek}
\usepackage{siunitx}
\usepackage{graphicx}

\title{Comparing simulations and test data of a radiation damaged charge-couple device for the Euclid mission} 

\author[a,*]{Jesper Skottfelt}
\author[a]{David Hall}
\author[a]{Jason Gow}
\author[a]{Neil Murray}
\author[a]{Andrew Holland}
\author[b]{Thibaut Prod'homme}
\affil[a]{Centre for Electronic Imaging, Dept. of Physical Sciences, The Open University, Milton Keynes MK7 6AA, UK}
\affil[b]{European Space Agency, ESTEC, Keplerlaan 1, 2200AG Noordwijk, The Netherlands}

\cftpagenumbersoff{figure}
\cftpagenumbersoff{table}

\begin{document} 
\maketitle
\begin{abstract}
The VIS instrument on board the Euclid mission is a weak-lensing experiment that depends on very precise shape measurements of distant galaxies obtained by a large CCD array. Due to the harsh radiative environment outside the Earth’s atmosphere, it is anticipated that the CCDs over the mission lifetime will be degraded to an extent that these measurements will only be possible through the correction of radiation damage effects.
We have therefore created a Monte Carlo model that simulates the physical processes taking place when transferring signal through a radiation-damaged CCD. 
The software is based on Shockley-Read-Hall theory, and is made to mimic the physical properties in the CCD as closely as possible. The code runs on a single electrode level and takes three dimensional trap position, potential structure of the pixel, and multi-level clocking into account. 
A key element of the model is that it also takes device specific simulations of electron density as a direct input, thereby avoiding to make any analytical assumptions about the size and density of the charge cloud.
This paper illustrates how test data and simulated data can be compared in order to further our understanding of the positions and properties of the individual radiation-induced traps.

\end{abstract}


{\noindent \footnotesize\textbf{*}Jesper Skottfelt,  \linkable{jesper.skottfelt@open.ac.uk} }\\

\begin{spacing}{1} 

\section{Introduction}
Radiation damage in detectors is an issue for most space mission. Outside the Earth's protective atmosphere a high flux of highly energetic particles will reach the detector array even if it is shielded by deliberate shielding materials, electronics and the spacecraft structure. 
In a radiative environment traps can be induced in the silicon lattice of a CCD. These traps are able to capture electrons from one charge package and release them into another at a later time, thus deteriorating the Charge Transfer Efficiency (CTE). This leads to a smearing of the image which can have a large impact on instrument performance. 
As higher and higher precision of the positional, photometric and shape measurements is required for current and future missions, it is therefore vital that radiation damage effects are corrected for with very high precision. 

An example of this is \textit{Euclid} \cite{Euclid_Red_book}, the second medium-class mission in the Cosmic Vision programme of the European Space Agency (ESA).
The Euclid spacecraft will orbit L2 in a large amplitude halo orbit, and it will therefore be subject to a relatively benign radiation environment compared to an Earth orbit. The radiation will mainly consist of solar particle events and Galactic cosmic rays \cite{Kohley2014}.
The scientific aim of the Euclid mission is to map the geometry of the Dark Universe using two instruments; the Visible Imager (VIS) \cite{VIS_SPIE_2012,VIS_SPIE_2014} and the Near Infrared Photometer Spectrometer (NISP) \cite{NISP_SPIE_2014}.
The VIS instrument is a large-scale imager, with a focal plane of 36 4K$\times$4K CCDs, that will do observations to enable Weak Lensing measurements. 
By measuring the ellipticity of the galaxies in most of the extra-galactic sky, it is possible to infer the mass distribution of the matter that distorts the galaxy shapes and thereby map the Dark Matter. 
In order for this experiment to be successful, the point spread function has to be very stable and tightly controlled and a very deep understanding of the systematic effects, especially the radiation damage effects, is therefore needed.

For this purpose we have created an Open University Monte Carlo model (OUMC) that can simulate charge transfer in radiation damaged CCDs.
The model is based on Shockley-Read-Hall theory \cite{Shockley_Read_1952, Hall_1952}, and the charge transfer is done on a single electrode level, building upon the heritage of a previous model iteration \cite{Hall_model_2012,Clarke_JINST_2012,Hall_optimisation_2012,Clarke_SPIE_2013}.
As opposed to most of the other radiation transfers codes that have been published, OUMC takes device specific charge distribution simulations as a direct input, thus eliminating the simplifications of an analytic solution for the charge distribution. This also means that subtleties such as multi-level clocking\cite{Murray_MultiLevel_2013} can be included in the model.
Although the simulation code is made for the Euclid VIS CCDs and how they will be operated, it is intended to be versatile so it easily can be set up to match other CCD architectures and operating conditions.

\section{Modelling Radiation Damage}
In the radiative environment outside the Earth's atmosphere, a CCD is subject to a large flux of high energy protons. 
These protons are able to displace atoms in the silicon lattice and thereby produce traps as detailed in Ref.~\citenum{Srour_2003}.
When the CCD is read out, the traps can capture electrons from one charge package and release them into a subsequent charge package and this leads to a smearing of the image. 

The capture and emission of electrons is described as decay processes in Shockley-Read-Hall theory. This means that they can be modelled using two exponential time constants; the capture time constant

\begin{equation} \label{eqn:tau_c}
\tau_c = \frac{1}{\sigma n v_{th}} 
\end{equation}
and the emission time constant
\begin{equation}
\tau_e = \frac{1}{X \chi \sigma N_c v_{th}}\exp\left(\frac{E}{kT}\right). 
\end{equation}
Here $\sigma$ is the capture cross-section, $n$ is the electron concentration, $v_{th}$ is the thermal velocity
\begin{equation}
v_{th} = \sqrt{\frac{3 k T}{m^{*}_{ce}}} \\
\end{equation}
$N_c$ is the density of states in the conduction band 
\begin{equation}
N_c = 2\left(\frac{2 \pi m^{*}_{de} k T}{h^2}\right)^{3/2} ,
\end{equation}
and $E$ is the energy level of the trap below the conduction band.
$m^{*}_{ce}$ and $m^{*}_{de}$ are the electron masses used for conductivity and density of states calculations, respectively.
$X$ is the entropy factor that is associated with the entropy change for electron emission and $\chi$ is a factor added to allow for any field enhanced emission that can affect the trap emission time as well as dark current generation \cite{Hopkinson_2004}.
The probability of a capture or emission of an electron over a given time $t$ can be calculated as 
\begin{equation} \label{eqn:Px}
P_x = 1 - \exp\left(\frac{-t}{\tau_x}\right) ,
\end{equation}
where $x$ can be substituted with $c$ or $e$ for capture and emission, respectively.

\begin{figure} [ht]
   \begin{center}
   \begin{tabular}{c} 
   \includegraphics[width=0.8\textwidth]{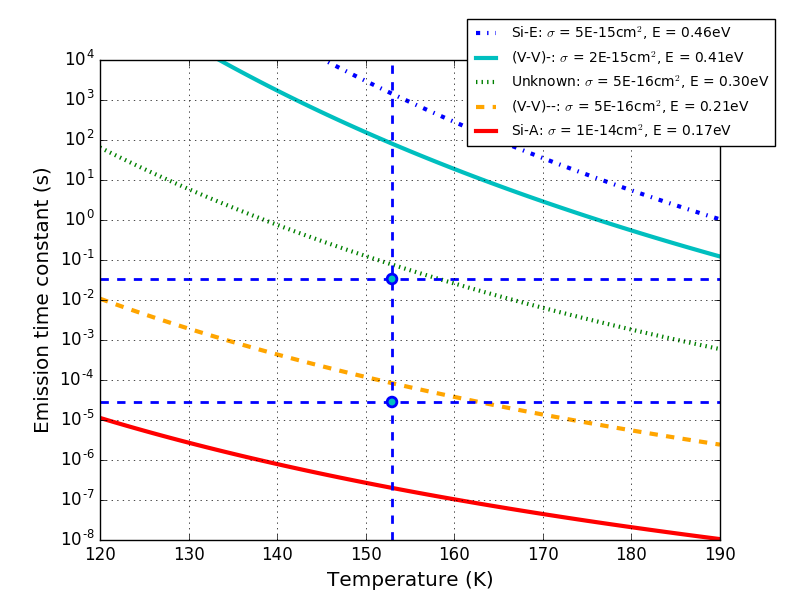}
   \end{tabular}
   \end{center}
   \caption[Emission time constants of well-known defects]{Emission time constants of different well-known defects as a function of temperature. The vertical dashed line indicates the nominal operating temperature (153 Kelvin) of the VIS detector array, and the horizontal dashed lines marks the parallel dwell ($t_{dwell}$) and transfer ($t_{shift}$) times as used in the experimental data acquisition and simulation describes in Sec.~\ref{sec:sim_vs_data}. \label{fig:tauE_temp} 
}
   \end{figure}

For n-channel CCDs there are a number of well-known defects with emission time constants in the range of typical integration and readout duration \cite{Hall_SPIE_2016}. 
These are plotted as a function of temperature in Fig.~\ref{fig:tauE_temp}. 
On the plot is also indicated the nominal operating temperature of the VIS detector array, and the two timings relevant for the simulation results described in Sec.~\ref{sec:sim_vs_data}.

As can be seen in Fig.~\ref{fig:tauE_temp}, there are some trap species with emission times much shorter than the parallel dwell and transfer times used in the simulations in Sec.~\ref{sec:sim_vs_data}. 
If the capture time constant is equally short for a given trap, then this trap is able to capture and emit many times within the timescales of interest. 
Eqn.~\ref{eqn:Px}, however, only takes a single capture or emission into account. 
It is therefore advantageous to define the combined probability of capture/emission of an electron by an empty/occupied trap after a time $t$ given both the capture and emission constants. 
Following the calculations made in Ref.~\citenum{Lindegren_trapping_Gaia,Thibaut_MC_2011}, the probability of a capture by an empty trap be expressed as
\begin{equation}
P_c = \frac{r_c}{r_{tot}}\left[1-\exp(-r_{tot}t)\right]
\end{equation}
and the probability of an emission by an occupied trap as
\begin{equation}
P_e = \frac{r_e}{r_{tot}}\left[1-\exp(-r_{tot}t)\right] ,
\end{equation}
where $r_x = \frac{1}{\tau_x}$ and $r_{tot} = r_c + r_e $.

\section{Charge Density Simulations}

\subsection{Electron Concentration}
Whereas the emission time constants can be found with high precision using techniques such as trap-pumping (cf. pocket-pumping \cite{Janesick_2001}, trap-pumping for trap location and efficiency \cite{Kohley_Gaia_2009,Mostek_2010,Murray_TrapPumping_2012}, trap-pumping for trap emission time constants \cite{Hall_trappumping_2014}), it is more difficult to estimate the capture time constant $\tau_c$ as this depends on the density distribution of the electron packet within the pixel.
This is highly dependant on pixel architecture and the nature and concentration of the dopants used in the manufacture process, and a precise analytical description of the charge density distribution $n$ is therefore difficult to obtain.

Several methods to circumvent this problem have been proposed over the years.
One approach is to use a $\beta$ parameter model\cite{Short_CDM_2013}, defined as
\begin{equation}
\frac{V_c}{V_g} = \left(\frac{N_e}{FWC}\right)^{\beta}. \label{eqn:beta_model}
\end{equation}
Here $V_c$ and $V_g$ is the volume of the charge cloud at the signal level $N_e$ and at Full Well Capacity $FWC$, respectively. The $\beta$ parameter can then be used to tune the confinement volume and can take values between 0 and 1. Effectively this gives 3 possibilities:
\begin{itemize}
\item $\beta=0$: The charge packet will fill the entire volume available to it no matter the size of the signal, thus only the density will change when the signal size increases (density-driven model).
\item $\beta=1$: The charge density will remain constant no matter the signal size, thus only the volume of the charge packet will increase as the signal increases (volume-driven model). 
\item $0<\beta<1$: Both the density and the volume of the charge packet will change with varying signal levels.
\end{itemize}

A similar approach has been used to mitigate the charge transfer inefficiency (CTI) effects on the ACIS camera on the Chandra X-ray Observatory. Here the $\beta$ parameter was measured experimentally to a value of $\sim 0.5$ \cite{Grant2003,Grant2004}.

Based on Silvaco ATLAS semiconductor software (see Sec. \ref{sec:silvaco}), a modified version of the $\beta$ parameter model was proposed by Refs.~\citenum{Hall_model_2012,Clarke_JINST_2012}:
\begin{equation}
V_c = \gamma N_e^{\beta} + \alpha,
\end{equation}
where $\gamma$ is a scaling constant. 
This model is better able to account for the contribution of the small signal using the parameter $\alpha$, leading to Eq.~\ref{eqn:beta_model} for small $\alpha$ or large $N_e$.

CTI correction for Hubble Space Telecope data is based on a fully volume-driven model\cite{Massey_Improved_2014}, where capture is assumed to be instantaneous within some volume defined by the signal level. 
Another approach is made in Ref.~\citenum{Thibaut_MC_2011}, where a density-driven model is used. This is done by modelling the density distribution as a normalised Gaussian function in three dimensions. 

However, common for these approaches is that the real physical solution is being fitted with an approximated analytical function, which will introduce some bias and which will only be suitable under certain operating conditions. 

\subsection{Silvaco Simulations} \label{sec:silvaco}
To mitigate this problem and to get as close to the actual physical properties as possible, we have introduced charge distribution simulation made for the specific device directly into the simulation.

The charge distribution simulations are made with Silvaco ATLAS semiconductor device simulation software \cite{Silvaco_ATLAS}. The ATLAS software can take a full 3D model of a pixel or register element of a CCD as input along with doping profiles, the temperature of the device and the voltages applied to the different phases of the CCD. 
This means that charge distribution in the device can be modelled under the exact operating conditions that is requested, and that the simulation can be redone if the operating conditions change. This includes the possibility for modelling the charge distribution when multi-level clocking etc. is applied. 

\begin{figure} [!ht]
   \begin{center}
   \begin{tabular}{c} 
   \includegraphics[width=\textwidth]{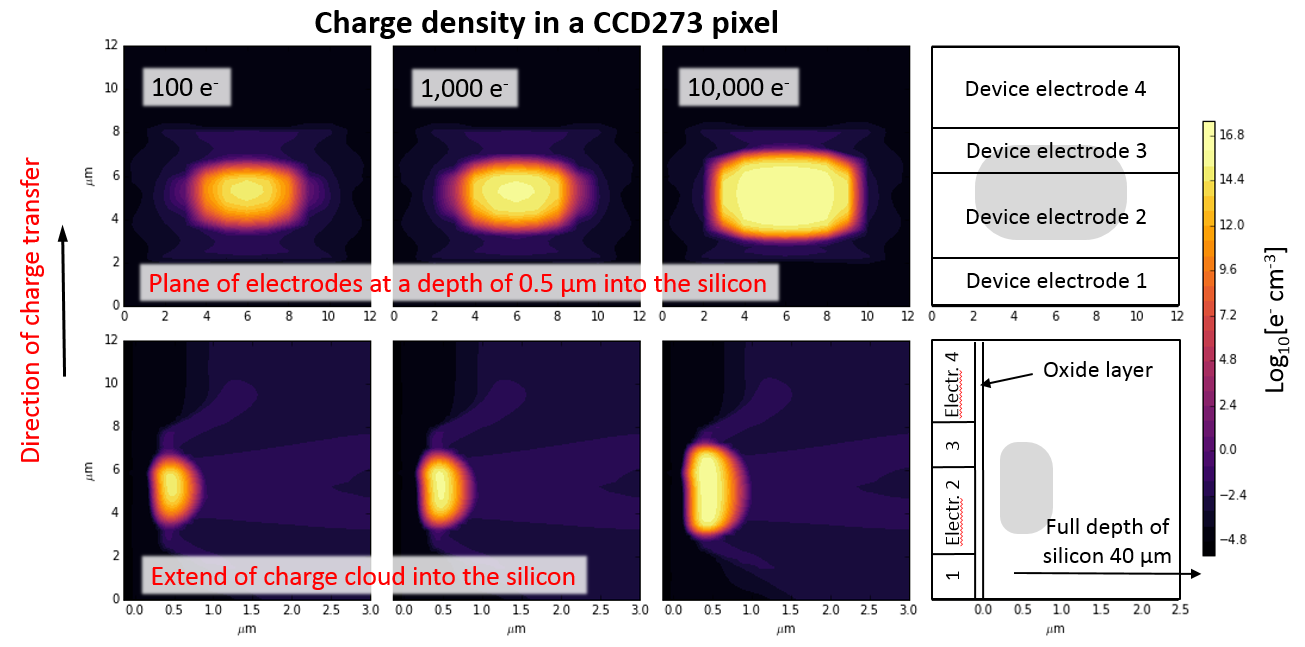}
   \end{tabular}
   \end{center}
   \caption[Charge density simulations]{ \label{fig:density_var_signal} 
2D cuts from 3D Silvaco ATLAS simulations of charge densities at different signal levels. The unit of the colorbar is $\log_{10}[\mathrm{electrons}/\mathrm{cm}^3]$. 
\textit{Upper row}: The charge cloud at the plane of the electrodes, with the electrodes aligned with the x-axis and at a depth into the pixel of 0.5~\textmu m. The plots shows the full 12x12 \textmu m pixel. 
\textit{Lower row:} The extend of the charge cloud into the pixel along the x-axis. The y-axis is the same as for the upper row, while the x-axis is only 3 \textmu m. }
   \end{figure}

The initial ATLAS modelling for the Euclid CCD273 pixel and serial register element is presented in Ref.~\citenum{Clarke_JINST_2012,Clarke_SPIE_2012}. 
These simulations have been redone using the current operating voltages and temperature as baselined for the VIS instrument. 2D cuts of the 3D charge density simulations are shown in Fig. \ref{fig:density_var_signal} in a CCD273 pixel. Each column shows a different signal level. 
The upper row shows the pixel in the plane of the electrodes at a depth of 0.5 \textmu m into the silicon. 
In the simulation the electrons are collected under phase 2 and 3 and the 4-2-4-2~\textmu m structure of the four phase pixel is evident.
The lower row shows the extend of the charge cloud into the silicon, and it is evident that most of the charge is collected in a buried channel close to the electrodes. 

\begin{figure} [!ht]
   \begin{center}
   \includegraphics[width=\linewidth]{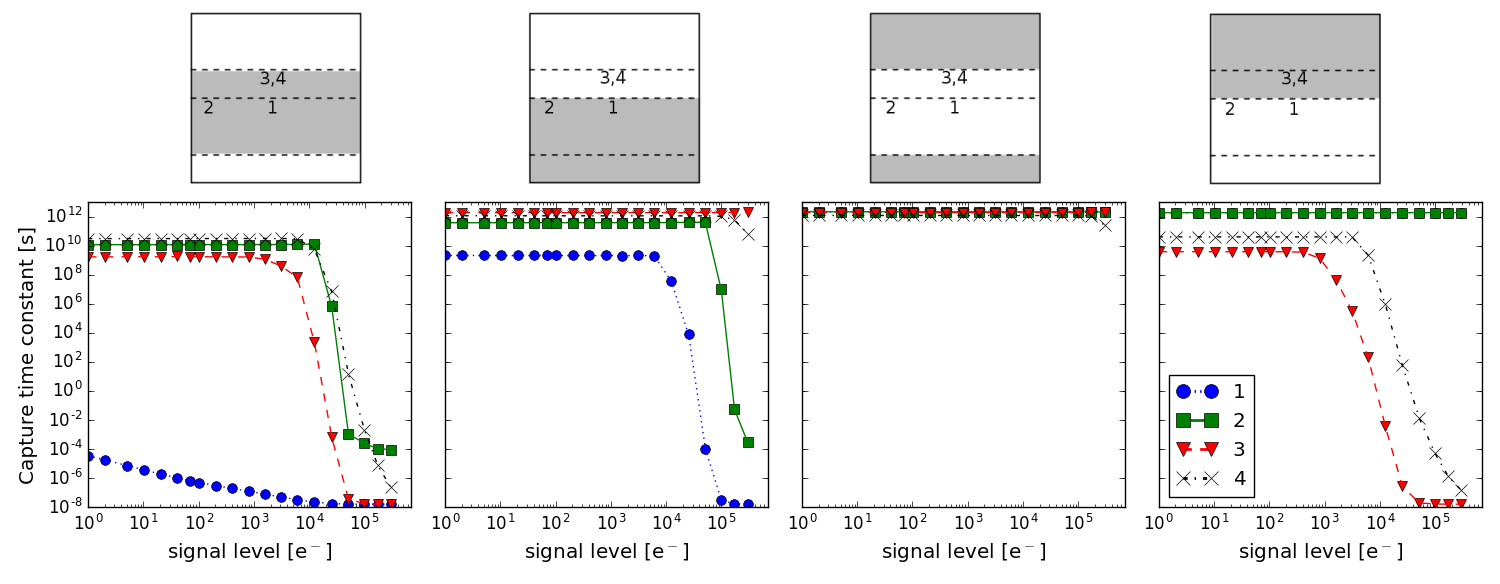}
   \end{center}
   \caption[Calculated capture time constants]{ \label{fig:tau_c} 
Calculated capture time constants $\tau_c$ for four traps at different positions. In each of the four columns, the charge has been shifted a single phase to simulate the readout of the CCD. 
\textit{Upper row:} The panels outline the positions of the four phases. The phases containing the charge are marked in grey, and the first column is thus similar to the situation in Fig.~\ref{fig:density_var_signal}. The numbers indicate the positions of four traps in the plane of the electrodes. Trap 1-3 is at a depth of 0.5~\textmu m into the silicon, and trap 4 is at a depth of 0.75~\textmu m.
\textit{Lower row:} The calculated capture time constants for the four traps using the trap positions and phases as shown in the panel above. 
}
   \end{figure}

The charge density simulations can then be used directly to calculate the capture time constants at the different signal levels as shown in Fig.~\ref{fig:tau_c}. 
This figure shows that there can be up to 20 magnitude difference in $\tau_c$ depending on the exact position of the trap in the pixel and that getting a precise value for the electron density therefore is important.

\subsection{Pixel Potential Structure}\label{sec:pix_pot}
Another issue where the Silvaco models can be used, is to determine which charge packet an emitted electron will join. 
As an electron will always move towards higher potential, it is the potential structure in the pixel that determines in which direction the electron will go. 

If a two-level clocking scheme is used and the phases have the same width, the potential structure can in most cases be estimated using symmetri considerations. 
However, in more complex situations, for instance if multi-level clocking is used and/or the phases in the pixels have varying width, it can be useful to extract the potential model from the Silvaco simulations.

   \begin{figure} [ht]
   \begin{center}
   \includegraphics[width=0.95\linewidth]{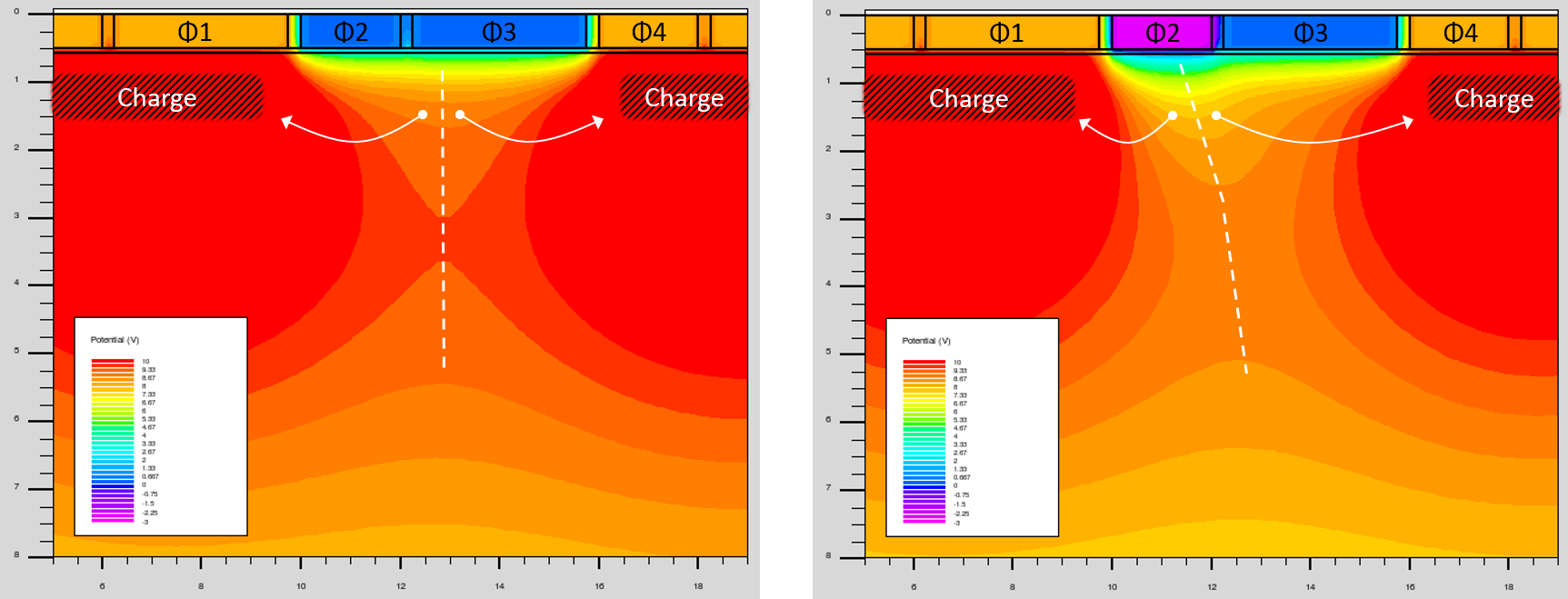}
   \end{center}
   \caption[Potential structure simulation]{ \label{fig:pot_model} 
   Silvaco simulations of the potential structure of the CCD273 pixel using two different voltage configurations; \textit{(left)} Normal two-level clocking with phases 2 and 3 at 0V, and \textit{(right)} tri-level clocking where phase 2 is at -3V and phase 3 is a 0V. Common for the two configurations is that phase 1 and 4 is at 8V. The upper limit of 10V is set to be able to see the contours of the potential well and the units on the axes are \textmu m. 
   The positions of the charge clouds and arrows showing the direction of the emitted electrons have been added to the plots.    }
   \end{figure} 

Figure \ref{fig:pot_model} shows the potential structure extracted from Silvaco simulations of two different voltage configurations for the CCD273 pixel. 
In the left panel a two-level clocking scheme is used, such that phases 1 and 4 are biased at 8V and phases 2 and 3 is at 0V. 
This means that the point of the lowest potential in the $x$-direction is located midway between the end of phase 1 and the start of phase 4, or $\sim$3~\textmu m from the rightmost edge of phase 1. Because of the 4-2-4-2~\textmu m structure of the pixel, the lowest potential is therefore not between phase 2 and 3, but $\sim$1~\textmu m into phase 3. 
An electron released from a trap at the leftmost fourth of phase 3 would therefore join the charge packet in phase 1, and not the charge packet in phase 4 as it would if the phases had been the same width. 

The right panel of Fig. \ref{fig:pot_model} shows an example of tri-level clocking. 
Here phases 1 and 4 are still biased at 8V and phase 3 at 0V, but phase 2 has been set to -3V.
The change of bias moves the point of the lowest potential some fraction of a phase width into phase 2, depending on the distance to the electrode.
This means that electrons released from all traps in phase 3 and traps in the rightmost part of phase 2 will join the charge packet in phase 4. 

Including this information in the model makes it possible to improve the precision of the charge transfer simulations, especially in the more complicated cases.

\section{Comparing Experimental and Simulated Data} \label{sec:sim_vs_data}
\subsection{Finding Emission Time Constants by Fitting to Charge Tails}

A common way of retrieving information about the traps in a CCD is from a charge tail, or Extended-Pixel-Edge-Response (EPER) tail.
A charge tail can be obtained from flatfield data, where the entire device is illuminated to a certain signal level. By reading out more pixels in the parallel (or serial) direction than there are in the array, called parallel (or serial) overscan pixels, a charge tail can be extracted.

\begin{figure} [!ht]
   \begin{center}
   \begin{tabular}{c} 
   \includegraphics[width=0.9\textwidth]{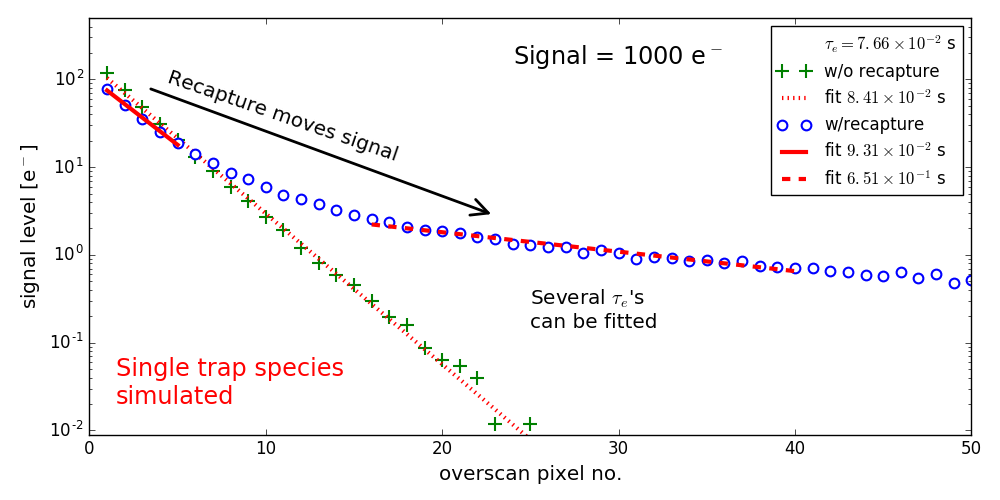}
   \end{tabular}
   \end{center}
   \caption[Charge tails with and without recapture]{ \label{fig:recap} 
   Charge tails from simulations of a single trap species with emission time constant of $\tau_e=7.66\times10^{-2}\,\mathrm{s}$ with and without recapture possible. A single exponential function are fitted to the charge tails to show that a single $\tau_e$ can be fitted very close to the real $\tau_e$ when recapture is not possible, while several $\tau_e$'s can be fitted in different regions, when recapture is possible. }
   \end{figure}

An approximation that is often made, is that the charge tail will only contain emitted electrons originating from the illuminated region. This means that the emission time constants of the traps in the array can be fitted directly with a sum of exponentials. 
In reality, however, recapture will occur from the charge tail itself, which will push charge from the beginning of the tail further down. 
An example of this can be seen in Fig.~\ref{fig:recap}, where a single trap species has been simulated both with and without recapture in the tail itself\footnote{Although this paper is based on Ref.~\citenum{Skottfelt2016}, updates to the simulation code, such as the inclusion of the pixel potential structure simulations as described in Sec.~\ref{sec:pix_pot}, have slightly improved some of the data and the plots that are based on these data (Figs.~\ref{fig:recap}, \ref{fig:trap_f1},\ref{fig:trap_f2} and \ref{fig:trap_f3}) are therefore slightly different.}. 
The simulations without recapture have been done by only allowing capture from pixels containing at least 90\% of the defined signal level.
When recapture is not possible, then it is easy to fit a single $\tau_e$, which is very close to the original $\tau_e$. 
However, in the more physically correct version where recapture from the tail is possible, then it is possible to fit multiple $\tau_e$'s making it difficult to retrieve the original $\tau_e$.
Furthermore if one is not aware of the recapture issue, there is a risk that instead of finding the right species, one will find several species with longer $\tau_e$'s than that of the correct trap. 

As fitting $\tau_e$'s to charge tails is instrumental in the current CTI correction algorithms (see Ref.~\citenum{Massey_Improved_2014,Israel_CTI_2015}) this issue could potentially have a large effect on the precision of the CTI correction if not taken into account. 

\subsection{Initial Results from the Open University Monte Carlo Model}
Even though the OUMC is still in an early version we demonstrate here how it can be used to estimate the types and densities of trap species in an irradiated device. 
For this purpose we are using flatfield data made as part of the testing campaign for the CCDs for the VIS instrument and detailed in Ref.~\citenum{Thibaut_SPIE_2014}. 

The method presented here is a process in several steps.
The first step is to simulate the individual trap species one at a time, but with all other parameters as close to the experimental setup as possible. 
In this case the flat fields were made at a number of different signal levels, and we have chosen  four of these, i.e. $N_e=[130,1000,8000,62000]\,\mathrm{e}^-$. The device is run at a temperature of 153 K 
and a single parallel pixel shift consist of four steps, which means that the four phase times used for each pixel shift in the parallel read out are 
$t_{ph}=[t_{dwell}+t_{shift},t_{shift},t_{shift},t_{shift}]$. Here $t_{dwell}=0.033\,\mathrm{s}$ is the time it takes to read out the serial register, and $t_{shift}=2.84\times10^{-5}\,\mathrm{s}$ is the time between each phase shift.
These phase times are indicated in Fig.~\ref{fig:tauE_temp} and we therefore choose to focus on the three middle species; the divacancies, (V-V)$^{-}$ and (V-V)$^{--}$ with energy levels of $0.41\,\mathrm{eV}$ and $0.21\,\mathrm{eV}$ respectively, and the ``unknown'' species for which the energy level is believed to be $\sim0.3\,\mathrm{eV}$.
The traps are not found to have discrete emission time constants, but rather a distribution of $\tau_e$'s. From trap pumping these distributions can be found experimentally (see Ref.~\citenum{Wood_TrapPumping_2014}) and substituted directly into the model.

Traps are currently deposited in random pixels and at random 3D-positions within the pixels, only based on the predefined trap densities. More research into the clustering of the traps in areas of the pixel\cite{Huhtinen2002} might improve the precision and will be considered for future versions of the model.

\begin{figure} [!ht]
   \begin{center}
   \begin{tabular}{c} 
   \includegraphics[width=0.9\textwidth]{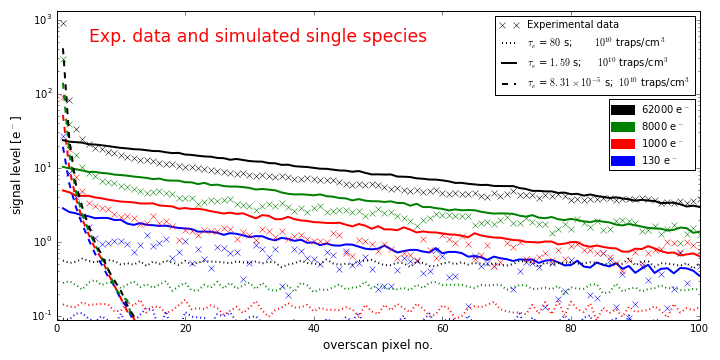}
   \end{tabular}
   \end{center}
   \caption[Charge tails for single species]{ \label{fig:trap_f1} 
   Charge tails from experimental data are shown as crosses with the four signal levels given as colors. Each trap species is simulated individually at each signal level using a fixed trap density of $10^{10}$ traps cm$^{-3}$ (First step). The resulting charge tails are represented by lines; dotted for (V-V)$^{-}$, dashed for (V-V)$^{--}$, and solid for the ``unknown'' species.
   }
   \end{figure}

\begin{figure} [!ht]
   \begin{center}
   \begin{tabular}{c} 
   \includegraphics[width=0.9\textwidth]{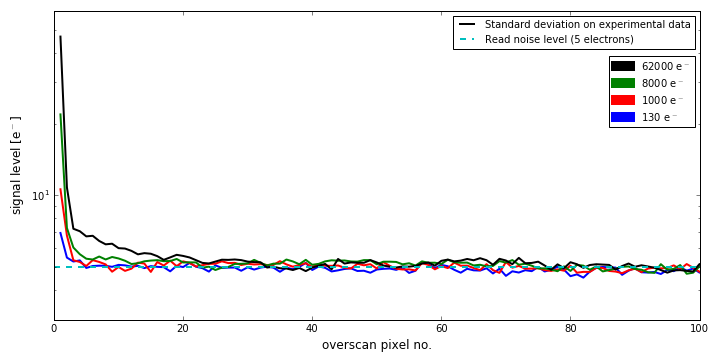}
   \end{tabular}
   \end{center}
   \caption[Uncertainty on experimental data]{ \label{fig:trap_stddev}
   The measured standard deviation over all the pixels in the charge tail, for each signal level. The standard deviation quickly centers around the read noise level of about 5 electrons (cyan dashed line) as expected. }
   \end{figure}

\begin{figure} [!ht]
   \begin{center}
   \begin{tabular}{c} 
   \includegraphics[width=0.9\textwidth]{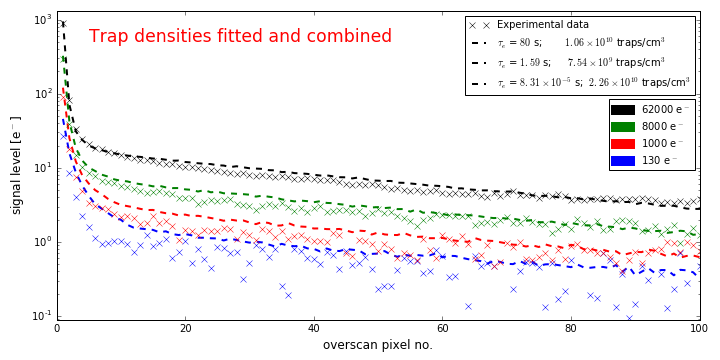}
   \end{tabular}
   \end{center}
   \caption[Charge tails for single species fitted and combined]{ \label{fig:trap_f2} 
   Charge tails from experimental data (crosses) same as in Fig.~\ref{fig:trap_f1}. 
For each signal level, the simulated single species charge tails from Fig.~\ref{fig:trap_f1} combined and fitted to the experimental data, such that the same trap densities for each trap species are used for all signal levels (second step). 
The dashed lines thus represents the combined charge tails with $1.06\times10^{10}$ traps cm$^{-3}$ for (V-V)$^{-}$, $2.26\times10^{10}$ traps cm$^{-3}$ for (V-V)$^{--}$, and $7.54\times10^9$ traps cm$^{-3}$ for the ``unknown'' species.
   }
   \end{figure}

\begin{figure} [!ht]
   \begin{center}
   \begin{tabular}{c} 
   \includegraphics[width=0.9\textwidth]{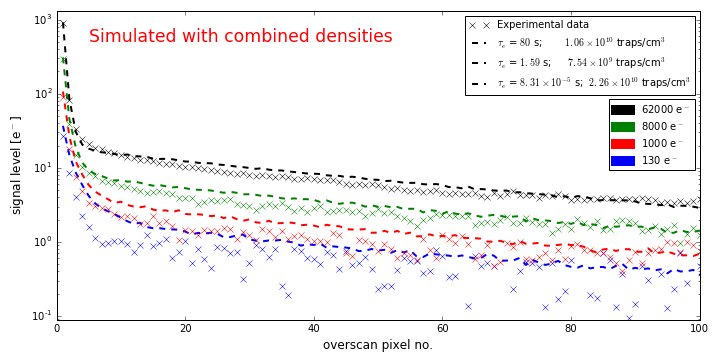}
   \end{tabular}
   \end{center}
   \caption[Charge tails simulated with found densities]{ \label{fig:trap_f3} 
   Charge tails from experimental data (crosses) same as in Fig.~\ref{fig:trap_f1}. For each signal level a simulation is run with all three trap species using the densities stated in Fig.~\ref{fig:trap_f2} (third step), and the resulting charge tails are represented by the dashed lines.    }
   \end{figure}

In Figs.~\ref{fig:trap_f1}, \ref{fig:trap_f2} and \ref{fig:trap_f3} the experimental charge tails at the different signal levels are plotted as crosses and are the same in all three figures as these are the ones that we try to match. 
Each experimental charge tail has been extracted from 20 images each containing 650 irradiated pixel columns, so each point in the charge tail is the mean value of $13,000$ pixel values.
As it is cumbersome to show errorbars on logarithmic plots, the errors on the experimental data is shown in Fig.~\ref{fig:trap_stddev}. The errors are calculated as the standard deviation over all $13,000$ pixel values for each overscan pixel, however it is clear that the read noise of about $5\,\mathrm{e}^-$ very quickly becomes the dominant component.

In Fig.~\ref{fig:trap_f1} the single species at a trap density of $10^{10} \,\mathrm{traps}\,\mathrm{cm}^{-3}$ have been over-plotted as lines. We find that the ``unknown'' species needs to be at $\sim0.34\,\mathrm{eV}$ to match the data. 
Note that each single trap species cannot be fitted with a single exponential, but needs a sum of several exponentials as described in section 4.1.

The second step in the process is to fit the experimental charge tails at each signal level with the combined tail of the three simulated species. 
The fitting parameter here is the trap density, and as the experimental data is made with the same device, it can be assumed that the trap density is the same for the four signal levels.
The best fit is found by minimising the sum of the $\chi^2$ for the each of the signal levels $N_e$, i.e. the value to minimise is
\begin{equation}
\sum_{N_e} \chi^2_{N_e} = \sum_{N_e} \sum^{M-1}_{i=0} \frac{\left[ D_{N_e}(x_i) - S_{N_e}(x_i) \right]^2 }{\sigma_{N_e}^2} ,
\end{equation}
where $D$ is the experimental data, $S$ is the simulated data, $\sigma$ is the noise on the experimental data, and $M$ is the total number of data points.
In Fig.~\ref{fig:trap_f2} the combined simulated charge tails using the best fit for the trap densities are shown. 

\begin{figure} [!ht]
   \begin{center}
   \begin{tabular}{c} 
   \includegraphics[width=0.9\textwidth]{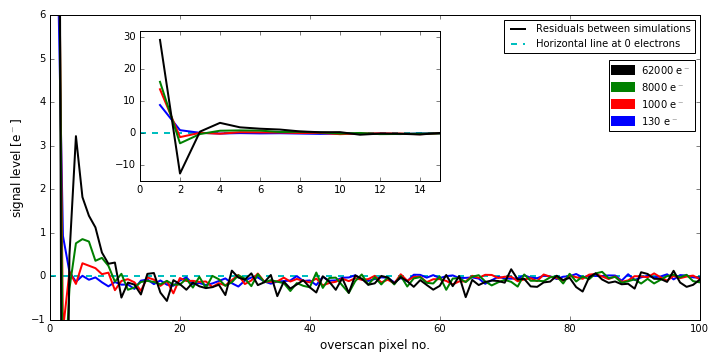}
   \end{tabular}
   \end{center}
   \caption[Residuals between combined and simulated charge tails]{ \label{fig:trap_residuals} 
   Residuals between the charge tails for single species fitted and combined (Fig.~\ref{fig:trap_f2}) and the charge tails simulated with found densities (Fig.~\ref{fig:trap_f3}. To better show the first part of the overscan pixels, a plot where the x-axis is zoomed in, and the y-axis zoomed out, is set in. }
   \end{figure}

The third step in the process is then to run the simulation using all three species with the fitted trap densities. The result of this is shown in Fig.~\ref{fig:trap_f3}. 
The residuals between the simulated data in this figure and the combined charge tails in Fig.~\ref{fig:trap_f2} is plotted in Fig.~\ref{fig:trap_residuals}. 
It is evident that these two tails are slightly different, and this is due to a difference in recapture as the signal levels in the tail is different from the single species simulation.
  
Depending on the level of accuracy needed, a possible fourth step can then be to see if a better fit for the trap densities can be found by running the simulation for varying trap densities. This, however, is a very time-consuming process, as a full simulation needs to be run for each small iteration of the trap densities. 
This last step has also been performed for the data presented here, but no noticeable improvement was found.

We find that the OUMC model is able to reproduce the experimental data very well. 
For the lowest signal level the simulation seems to give a bit too high values, however, at such low signal levels the uncertainties coming from calibration errors, might be an issue.
As recapture is an intrinsic part of the model, and we are not dependant of fitting exponentials to the charge tails, we are able to make very precise estimates of the emission time constants, densities and other physical parameters of the traps. 

As part of the Euclid radiation damage study at Centre for Electronic Imaging at the Open University, a large amount of experimental data will be obtained from a number of CCD273 devices both pre and post irradiation.
These tests include trap-pumping and a number of different CTE measurements, all done at different temperature and signal levels. 
These tests can be used directly to test the OUMC, and will be an important part of a further validation of the model.

\section{Conclusions}
A new implementation of the Open University Monte Carlo model (OUMC), for simulating charge transfer in a radiation damaged CCD, is presented. 
It is shown that the electron density has a large effect on the capture time constant and that it therefore can have a big influence on the precision of the simulation. 
Instead of making analytical assumptions on the size and density of the charge cloud, the OUMC therefore takes charge distribution simulations as a direct input. 

It is illustrated how the OUMC can be used to estimate the density and energy levels of the different trap species in an irradiated CCD.
By fitting simulated charge tails of single trap species to experimental data at different signal levels, we show that the experimental data can be reproduced with high precision. 
However, further validation of the model is needed and data for this will be obtained as part of the CCD273 radiation damage study done for the Euclid mission.

\bibliography{jatis2016} 
\bibliographystyle{spiejour} 

\end{spacing}
\end{document}